\documentclass[journal]{IEEEtran}
\usepackage[utf8]{inputenc} 
\usepackage[T1]{fontenc}    
\usepackage{hyperref}       
\usepackage{url}            
\usepackage{booktabs}       
\usepackage{amsfonts}       
\usepackage{nicefrac}       
\usepackage{microtype}      
\usepackage{graphicx}
\usepackage{amssymb,amsmath,amsthm,array,amsfonts}
\usepackage{xcolor}
\usepackage{booktabs}
\usepackage{underoverlap}
\usepackage{subfigure}
\usepackage{multirow} 
\usepackage[rightcaption]{sidecap}
\sidecaptionvpos{figure}{c}
\usepackage{balance}
\usepackage{pifont}
\balance

\newcommand{\cmark}{\ding{51}}%

\title{Snow avalanche segmentation in SAR images with Fully Convolutional Neural Networks}

\author{Filippo Maria Bianchi$^{*}$,
        Jakob Grahn,
        Markus Eckerstorfer,
        Eirik Malnes,
        Hannah Vickers
\thanks{*filippo.m.bianchi@uit.no}
\thanks{F. M. Bianchi is with the Dept. of Mathematics and Statistics, UiT the Arctic University of Norway and with NORCE, The Norwegian Research Centre AS}
\thanks{J. Grahn, M. Eckerstorfer, E. Malnes, H. Vickers are with NORCE, The Norwegian Research Centre AS}%
}

\begin{document}

\maketitle

\begin{abstract}
Knowledge about frequency and location of snow avalanche activity is essential for forecasting and mapping of snow avalanche hazard. 
Traditional field monitoring of avalanche activity has limitations, especially when surveying large and remote areas. 
In recent years, avalanche detection in Sentinel-1 radar satellite imagery has been developed to improve monitoring. 
However, the current state-of-the-art detection algorithms, based on radar signal processing techniques, are still much less accurate than human experts.
To reduce this gap, we propose a deep learning architecture for detecting avalanches in Sentinel-1 radar images. 
We trained a neural network on $6,345$ manually labelled avalanches from $117$ Sentinel-1 images, each one consisting of six channels that include backscatter and topographical information. 
Then, we tested our trained model on a new SAR image. 
Comparing to the manual labelling (the gold standard), we achieved an F1 score above $66\%$, while the state-of-the-art detection algorithm sits at an F1 score of only $38\%$. 
A visual inspection of the results generated by our deep learning model shows that only small avalanches are undetected, while some avalanches that were originally not labelled by the human expert are discovered. 
\end{abstract}

\begin{IEEEkeywords}
Deep Learning; Saliency Segmentation; Convolutional Neural Networks; Snow Avalanches; SAR; Sentinel-1.
\end{IEEEkeywords}

\section{Introduction}
Knowledge about the spatio-temporal distribution of snow avalanche (hereafter referred to as avalanche) activity in a given region is critical for avalanche forecasting and hazard mapping. An increase in avalanche activity or magnitude of releasing avalanches leads to an increase in avalanche risk.
Conventionally, avalanche activity is primarily monitored through field measurements, which is time-consuming, expensive, and can only be done for very few accessible areas. 
Monitoring avalanche activity using satellite-borne synthetic aperture radar (SAR) has, therefore, gained considerable interest in recent years.
SAR products enable continuous covering of very large areas, regardless of light and weather conditions~\cite{eckerstorfer_complete_2017}.

An experienced operator can identify avalanche debris (the depositional part of an avalanche) in SAR change detection composites (showing temporal radar backscatter change) with high accuracy.
On the other hand, automatic signal processing methods based on radar backscatter thresholding and segmentation often fail and produce a large number of false alarms due to the highly dynamic nature of snow in the SAR images~\cite{vickerssynthetic}. 
A key limitation of classical segmentation methods is that they mainly focus only on pixel-wise information in radar backscatter, without accounting for the contextual information around the pixel and high-level features, such as the shape and the texture of avalanche debris.
Also, local topography in which the avalanches occur is largely disregarded since it has only been used to mask out areas where avalanches are unlikely to occur. 
However, the occurrence of avalanches is strongly correlated to topographical conditions and avalanche debris exhibits characteristic shapes, which should both be taken into account when performing the detection.

Convolutional neural networks (CNNs) have attracted considerable interest for their ability to model complex contextual information in images~\cite{zhu2017deep}. 
Prominent examples in remote sensing are terrain surface classification~\cite{zhou2016polarimetric, kampffmeyer2018urban}, categorization of aerial scenes~\cite{penatti2015deep}, detection of changes in the terrain over time from SAR and optical satellite sensors~\cite{8798991, 8517033}, and segmentation of objects from airborne images~\cite{kampffmeyer2016semantic, bianchi2020large}.
Nevertheless, few research efforts have been devoted so far towards detecting avalanche activity from SAR data, which remains an open and challenging endeavour. 
In our previous work~\cite{kummervoldavalanche}, we proposed a deep learning architecture to perform binary classification of avalanches in Northern Norway. 
In particular, we used a CNN to classify fixed-size patches of SAR images in two classes: 1 if the patch contains at least one avalanche, or 0 otherwise.
Our approach was successively adopted later on for SAR-borne avalanche detection in the Alps~\cite{sinha2019can} and in other locations in Norway~\cite{waldeland2018avalanche}. 
As a major limitation, patch-wise classification cannot determine the presence of multiple avalanches within the same patch.
Also, the results are heavily influenced by the patch size, which makes it difficult to evaluate the detection performance.
In particular, for large windows is easier to correctly predict the presence of at least one avalanche, but the resolution of the detection is too coarse and not very useful. 

In this work, we approach avalanche detection as a saliency segmentation task, where the classification is not done at the patch level, but rather at the individual pixel level. 
We adopt a Fully Convolutional Network (FCN) architecture, which generates for each input image a segmentation mask. 
This solves the drawback of the dependency from the window size and makes it possible to determine the exact location of the avalanches.
Our work provides important contributions to the fields of Earth science, remote sensing, and avalanche risk assessment. 
\begin{itemize}
    \item We explore, for the first time, the capability of deep learning models in detecting the presence of avalanches in SAR products at a pixel granularity and surpass the current state-of-the-art avalanche detection algorithm~\cite{vickerssynthetic}. 
    Our work was possible thanks to a large dataset of SAR products manually annotated by an avalanche expert. 
    \item We advance our knowledge on topographical features to identify areas where avalanches are highly likely to occur.
\end{itemize}

Notably, we introduce a new topographical feature, called \textit{potential angle of reach} (PAR), which indicates how likely it is for an avalanche to reach a specific location. 
We do not use the PAR to filter input images or detection results, but we rather provide the PAR as an exogenous input feature to the FCN.
We first estimate how informative is the PAR in the discrimination of avalanche and not-avalanche pixels. 
Then, in the experimental section, we evaluate how much the detection performance of the deep learning model improves when providing the FCN with the PAR feature map.

\section{SAR Dataset}

\begin{figure*}[!ht]
    \centering
    \subfigure[VV difference]{
        \includegraphics[width=6cm, height=6cm]{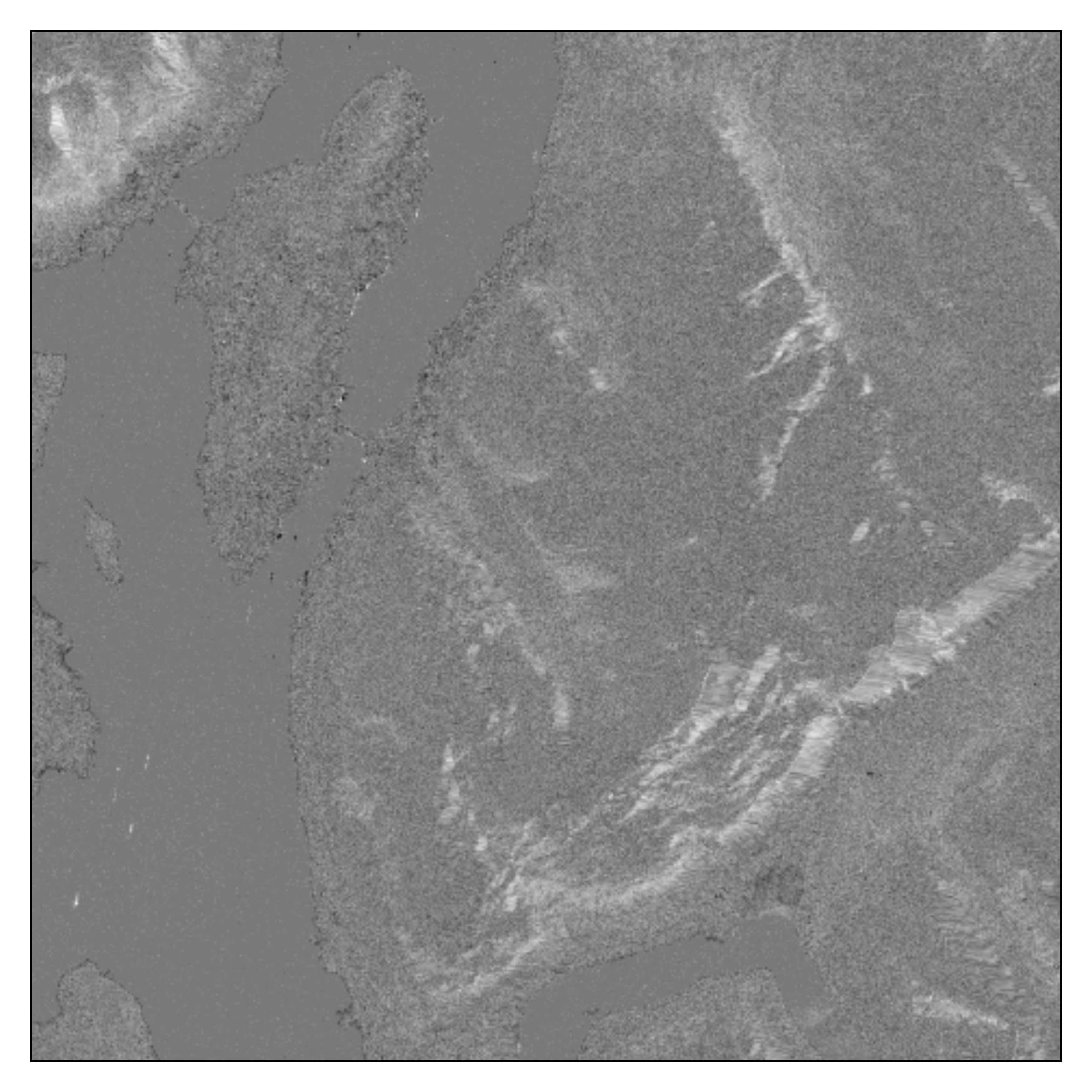}
    }
    \hspace{-.8cm}
    ~
    \subfigure[VH difference]{
        \includegraphics[width=6cm, height=6cm]{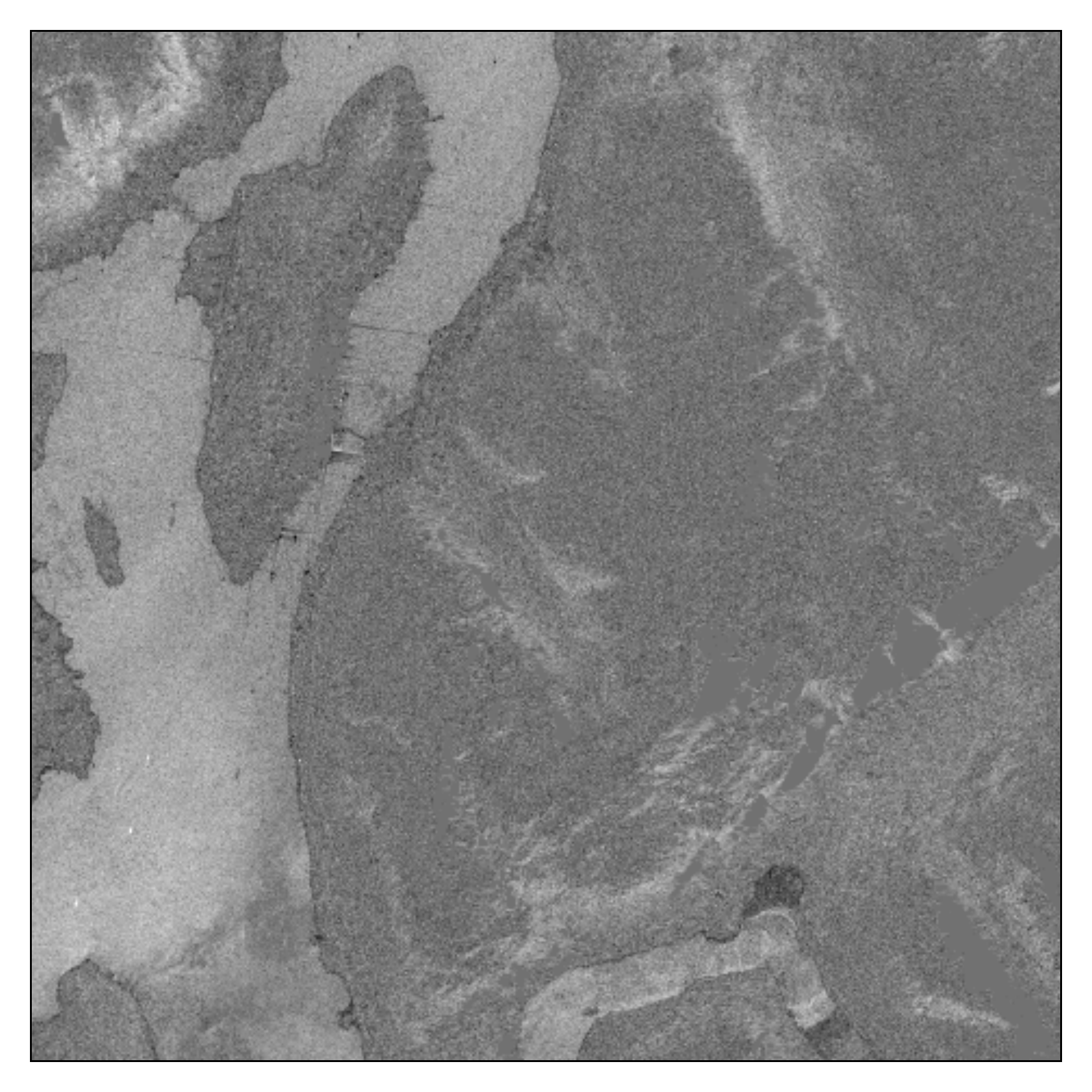}
    }
    \hspace{-.8cm}
    ~
    \subfigure[VVVH]{
        \includegraphics[width=6cm, height=6cm]{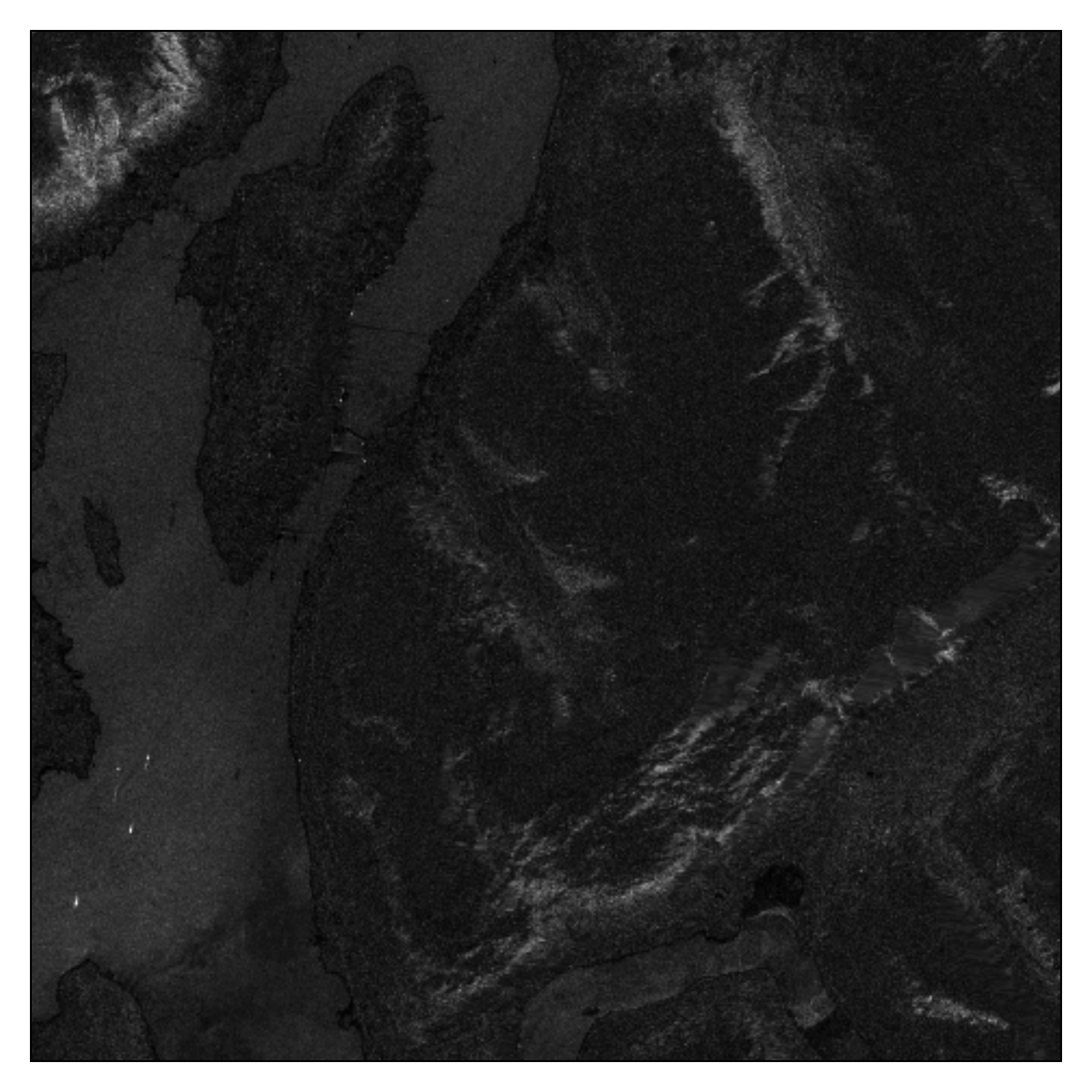}
    }
    
    \subfigure[Slope]{
        \includegraphics[width=7.5cm, height=6cm]{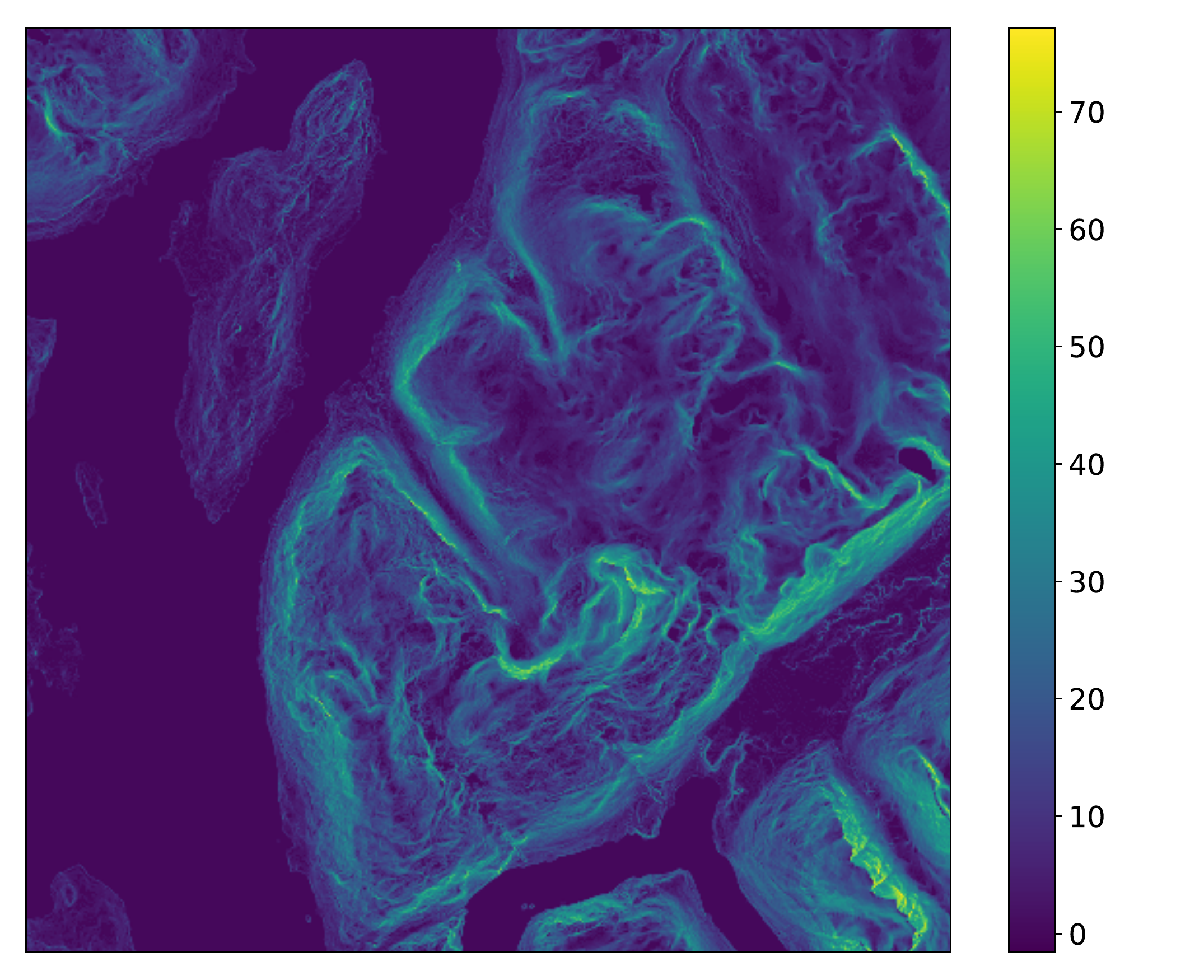}
    }
    ~
    \subfigure[PAR]{
        \includegraphics[width=7.3cm, height=6cm]{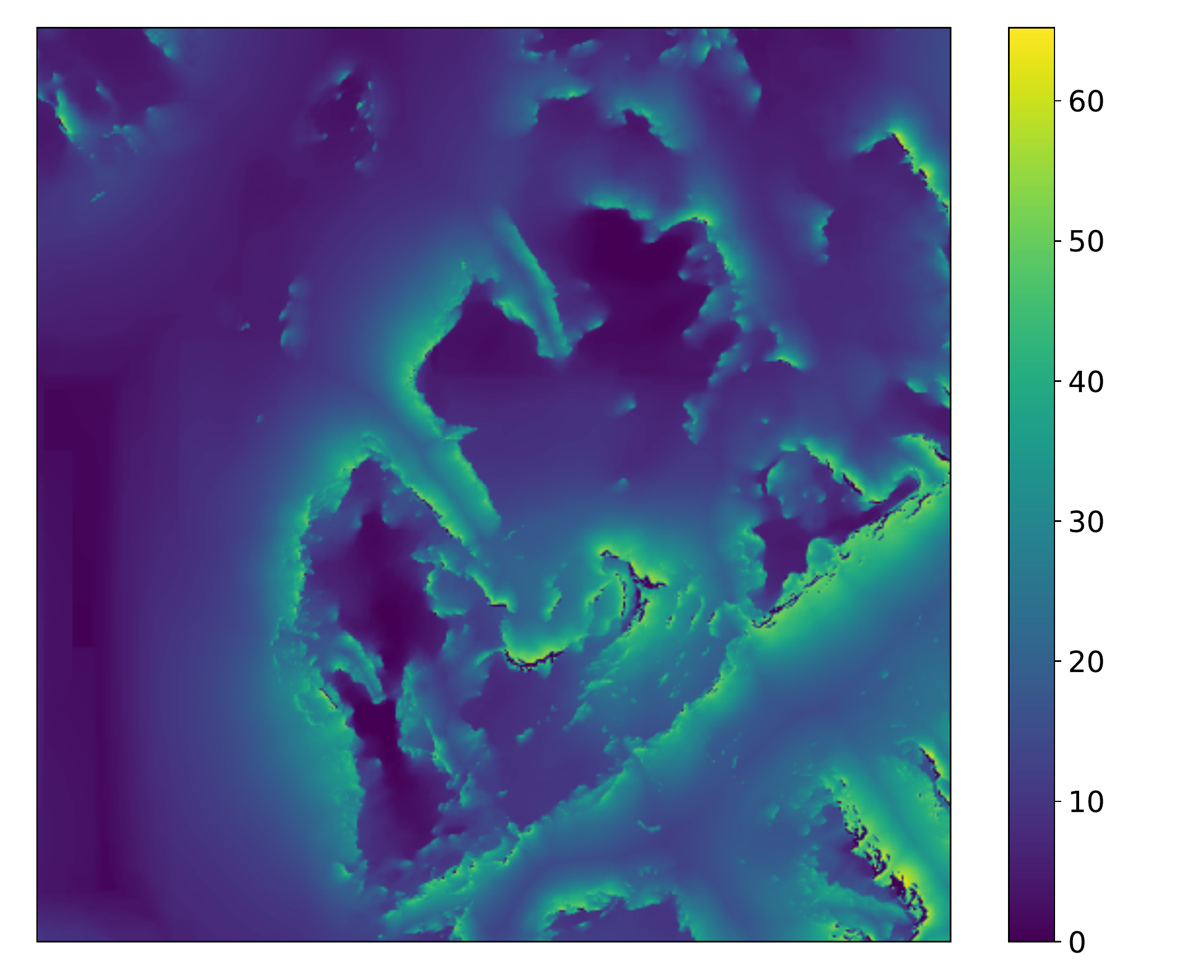}
    }
    \caption{(a, b) the SAR features obtained from the difference in the VV and VH channels. 
    (e) product VVVH of the squared differences. 
    (d, e) slope and PAR feature maps. 
    Only a small area ($1k \times 1k$ pixels) of the actual scene is depicted here.}
    \label{fig:sar_input}
\end{figure*}

The dataset consists of data from the Sentinel-1 (S1) satellites. In particular, data acquired in the interferometric wideswath (IW) mode was considered in terms of the ground range detected (GRD) product. In total, 118 SAR scenes covering two mountainous regions in Northern Norway in the period Oct. 2014-Apr. 2017. 

\subsection{Preprocessing}
Each SAR product was 
(i) radiometrically calibrated to radar backscatter (sigma nought) values, 
(ii) spatially downsampled from 10 to 20 meters resolution, 
(iii) geocoded onto a 20 meters resolution UTM-grid (EPSG:32633) using a 10 meter resolution digital elevation model (DEM) ~\cite{planesailing}, 
(iv) radiometrically transformed to decibel (dB) values and clipped to range values from -25 to -5 dB, to remove noise and restrict the range of the backscatter to intervals where avalanches are visible. 
The preprocessed products were then grouped by their satellite geometry, such that the scenes within a group have the same viewing geometry. For each group, scenes were paired chronologically into \textit{reference} and \textit{activity} image pairs. 
For the two S1 satellites, the reference image is acquired 6 days before the activity image (12 days before the launch of S1B in 2015).
The resulting products have an approximate size of $11.500 \times 5.500$ pixels, and each pixel covers $20 \times 20$ meters.

\subsection{Generation of SAR features}
We considered three SAR features to generate the images to be processed by the deep learning model.
The first two are the difference of the horizontal and vertical polarization between the reference and the activity image: VV = VV\textsuperscript{activ} - VV\textsuperscript{ref}, VH = VH\textsuperscript{activ} - VH\textsuperscript{ref}. 
The difference values are re-scaled to [0,1] (see Fig.~\ref{fig:sar_input}(a,b)). 
The third feature is the point-wise product of the difference images squared: VVVH = $\text{VV}^2 * \text{VH}^2$ (see Fig.~\ref{fig:sar_input}(c)).
We did not consider radar shadow, layover masks, or land masks depicting avalanche runout zones, which are not available for all areas. 

\subsection{Labeling}
For each product, a human expert generated a binary segmentation mask that indicates whether a pixel in the product is an avalanche or not. 
To create the segmentation mask, the human expert looked for changes in a difference image obtained from the following three channels: R[VV\textsuperscript{reference}], G[VV\textsuperscript{activity}], B[VV\textsuperscript{reference}].
We considered visual detection as the golden standard and used it as ground truth to train and evaluate our deep learning model.
The whole dataset contains a total of $6,345$ avalanches; $3,667,355,474$ pixels are classified as ``non-avalanche'' and $712,945$ ($0.000194\%$ of the total) as ``avalanche''.

\section{Topographical features}
Since avalanches are caused by steep terrain, the topography is an important factor to determine where avalanches can appear. In particular, the local slope needs to be steep enough for an avalanche to release and the slope typically needs to flatten out for the avalanche to stop. 
Therefore, it is reasonable to consider such information when performing the detection task and we generated two feature maps from the digital elevation model (DEM), which is available for the entire Norway in 10m pixel resolution. 
The first is the local slope angle of the terrain; the second is a new topographical feature introduced in this work, which is called \textit{potential angle of reach} (PAR).

\subsection{Slope angle} 
The slope angle feature map is directly computed by taking the gradient of the DEM (see Fig.~\ref{fig:sar_input}(d)).
The terrain slope is often considered when detecting avalanches, as they typically start in terrain between 35-45 degrees steepness and deposit on less steep slope angles.  
In previous work, the slope was used to derive a runout mask that indicated where avalanches are most likely to deposit~\cite{vickerssynthetic}. 
Since the mask is applied to filter out areas in a pre-processing operation, the slope feature did not contribute to the actual detection.
Most importantly, since run-out masks are obtained by manual thresholding the slope, if a wrong threshold is chosen some avalanches will not be detected.
To address this issue, we provide the slope as an additional layer of the input image and let our neural network learn how to optimally exploit it to solve the segmentation task, without applying manually chosen thresholds.

\begin{figure}[h!]
 \centering
 \includegraphics[width=0.8\columnwidth]{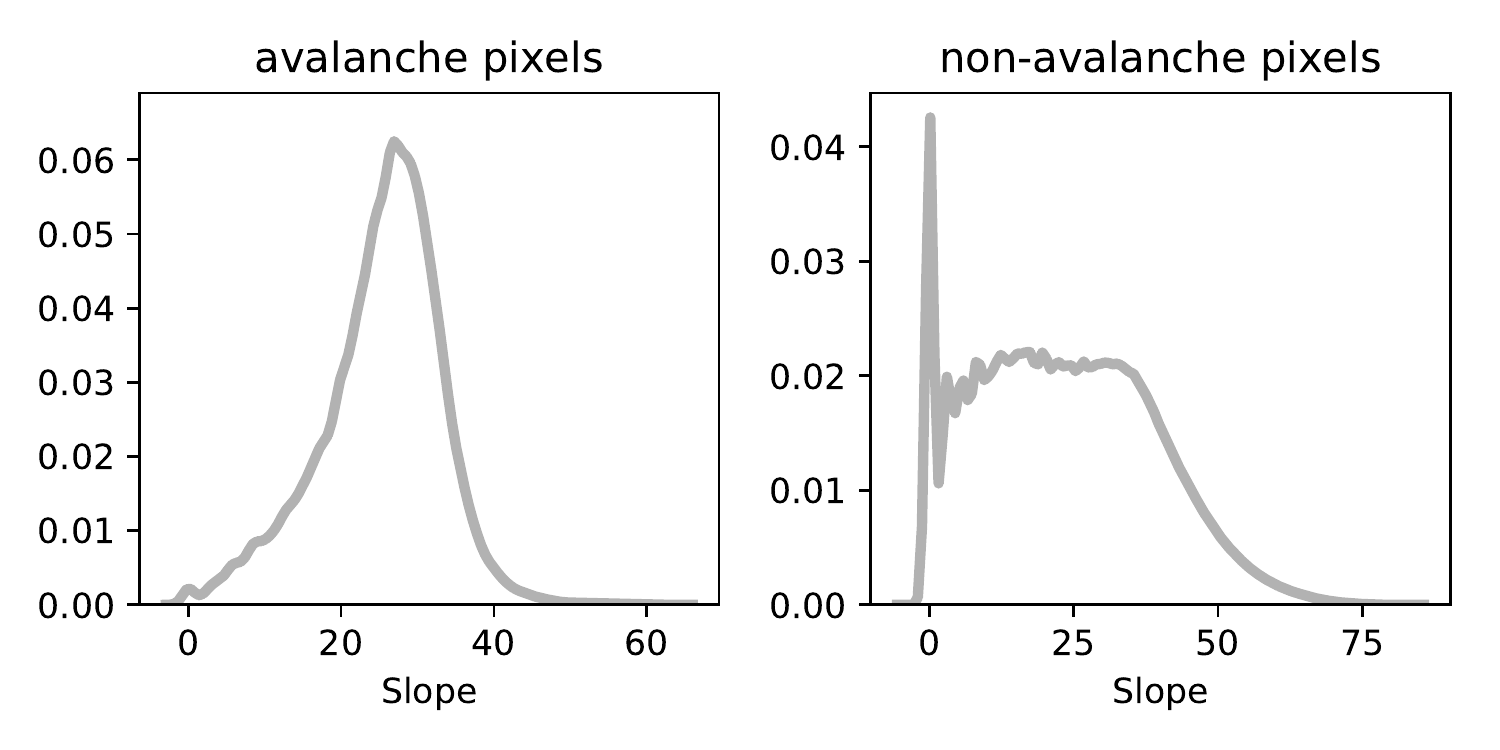}
 \vspace{-.5cm}
 \caption{Distribution of the slope angle for avalanche and non-avalanche pixels}
 \label{fig:d_slope}
\end{figure}

Fig.~\ref{fig:d_slope} shows that the distribution of the slope angle is different for the avalanche and non-avalanche pixels in our dataset.
In particular, avalanche pixels are mostly concentrated around [20, 35] degrees.
The difference in the two distributions indicates that the slope angle can be exploited to discriminate between ``avalanche'' and ``non-avalanche'' classes.

\subsection{Potential angle of reach (PAR)}
The angle of reach of an avalanche, sometimes denoted $\alpha$ and referred to as the alpha-angle, indicates how far an avalanche travels from its triggering point in relation to the descent it makes. 
Specifically, it is defined as the elevation angle of the line between the point of furthest avalanche runout and the point of highest release. 
For most avalanches, this angle ranges between 20 to 40 degrees~\cite{bakkehoi_domaas_lied_1983, delparte2008statistical, johnston2011estimating}. 

\begin{figure}[h!]
 \centering
 \includegraphics[width=0.35\columnwidth]{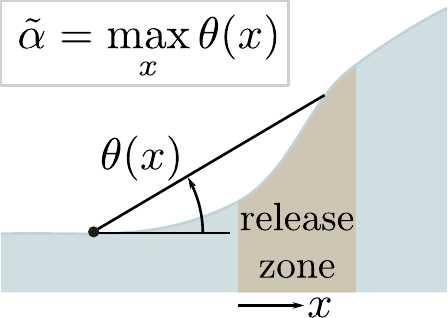}
 \vspace{-.4cm}
 \caption{Definition of the potential angle of reach $\tilde\alpha$, where $\theta(x)$ denotes the angle between the horizontal and the line drawn from a point in a release zone, denoted $x$, to the point of interest.}
 \label{fig:angle_of_reach}
\end{figure}

While the angle of reach is defined only for an existing avalanche, we here introduce the \emph{potential angle of reach} (denoted as $\tilde\alpha$), which is defined for a hypothetical avalanche located at any given point in the DEM. Ideally, this feature will range values between 20-40 degrees in terrain where avalanches can accumulate. 
Assuming that avalanches normally releases in steep terrain, \textit{e.g.}, in slopes of 30-50 degrees, the PAR angle is obtained by (i) computing the elevation angle to all neighbouring release points $x$ (within a 4 km radius), and (ii), by taking the maximum of all such angles, as illustrated in Fig.~\ref{fig:angle_of_reach}. 
By computing the PAR for each point in the DEM, a PAR feature map can be obtained and used as an additional channel of input images. 

\begin{figure}[h!]
 \centering
 \includegraphics[width=0.8\columnwidth]{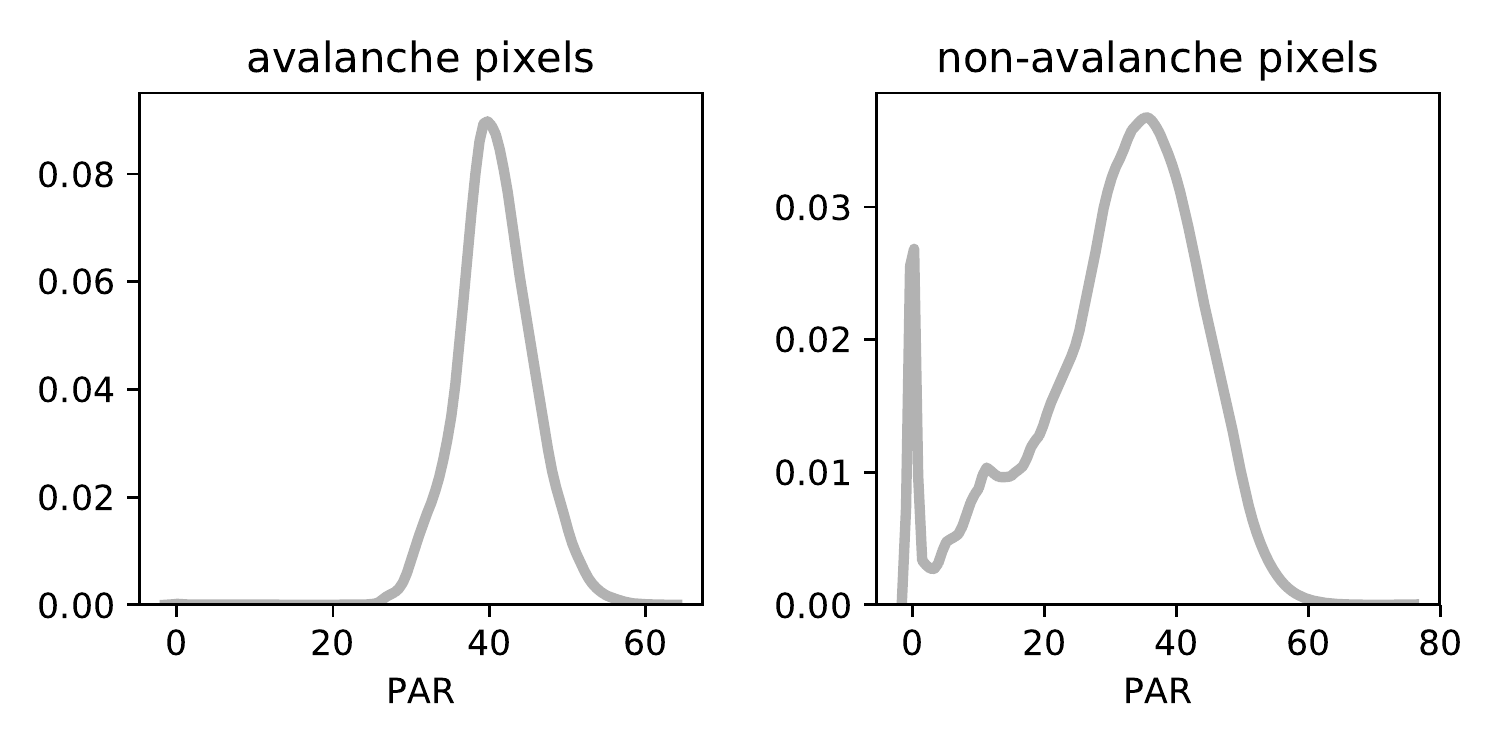}
 \vspace{-.5cm}
 \caption{Distribution of the PAR for avalanche and non-avalanche pixels}
 \label{fig:d_angle}
\end{figure}

Fig.~\ref{fig:d_angle} depicts the distribution of the PAR angles for avalanche and non-avalanche pixels using the training data. It is possible to see that for avalanche pixels the distribution is more regular and has a single peak centred around 40 degrees. While the true angle of reach is expected to range 20 to 40 degrees, the PAR is consequently biased towards higher values. 
We concluded that the PAR is informative since the two distributions are different for the two classes.
Contrarily to the slope, the PAR is not simply concatenated to the other layers of the input image but is rather used to encourage the deep learning model to focus on specific areas (see Sect.~\ref{sec:attention}).

\section{Deep Learning Model}

\begin{figure}[ht!]
 \centering
 \includegraphics[width=\columnwidth]{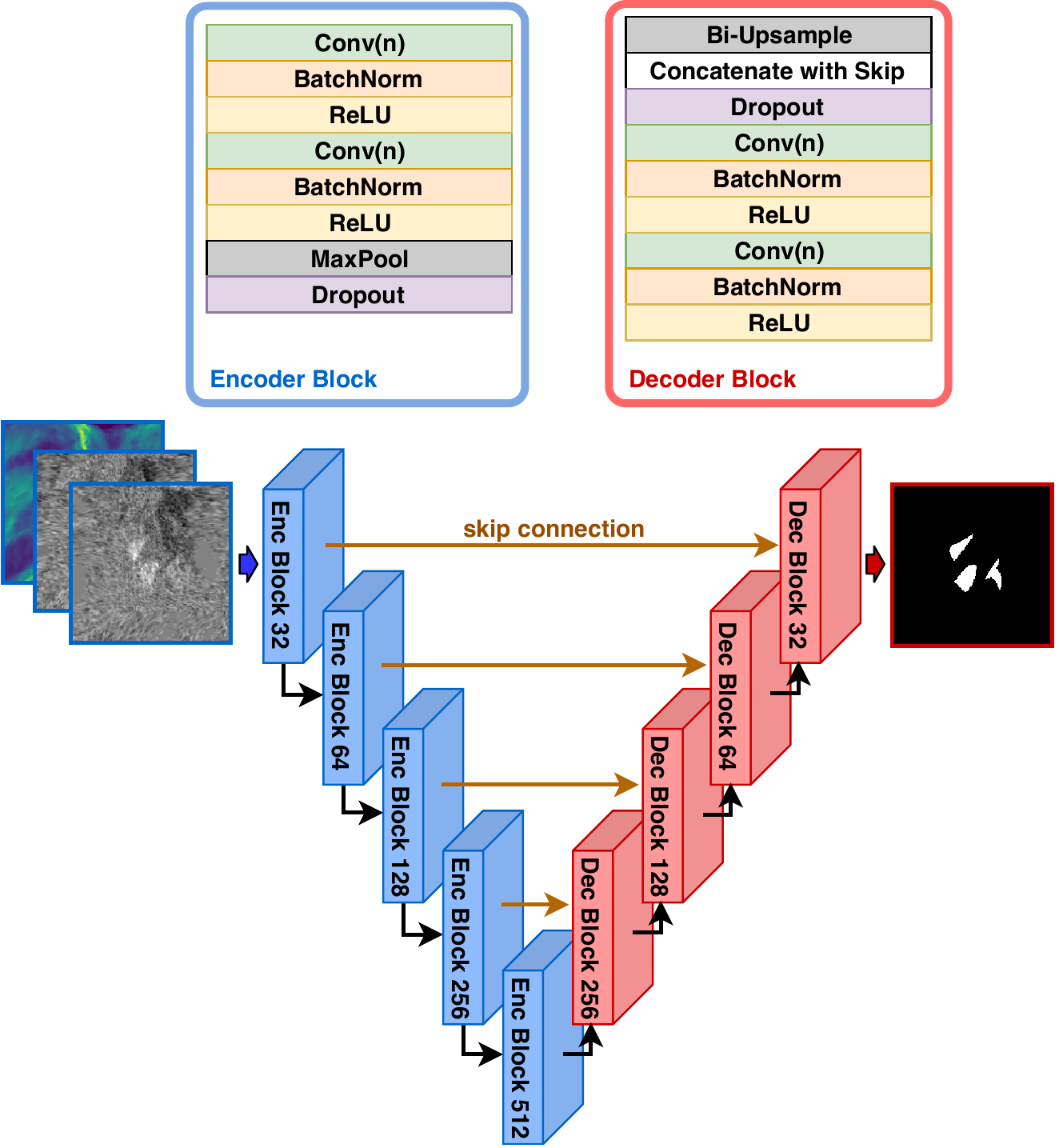}
 \caption{The FCN architecture used for segmentation. Conv($n$) stands for a convolutional layer with $n$ neurons. For example, $n=32$ in the first Encoder Block, $64$ in the second, and so on.}
 \label{fig:unet}
\end{figure}

The FCN network used for segmentation is based on the U-Net architecture~\cite{ronneberger2015u}, which consists of an \textit{encoder} and a \textit{decoder}, respectively depicted in blue and red in Fig.~\ref{fig:unet}.
The encoder hierarchically extracts feature maps that indicate the presence of the patterns of interest in the image.
By reducing the spatial dimensions and increasing the number of filters, the deeper layers in the encoder capture patterns of increasing complexity and with a larger spatial extent in the input image.
The decoder gradually transforms the high-level features and, in the end, maps them into the output.
The output is a binary segmentation mask, which has the same height/width of the input and indicates which are the pixels that belong to the avalanche class.
The skip connections link the feature maps from the encoding to the decoding layers, such that some information can bypass the bottleneck located at the bottom of the ``U'' shape.
In this way, the network still learns to generalize from the high-level latent representation but also recovers the spatial information through a pixel-wise semantic alignment with the intermediate representations.

Fig.~\ref{fig:unet} shows the architecture details: the number $n$ in each Enc/Dec Block indicates the quantity of $3 \times 3$ filters in the Conv($n$) layers.
The encoder reduces the spatial dimension with max pooling, while the decoder restores it through bilinear upsampling.
Each block contains 2 Batch Normalization~\cite{ioffe2015batch} and one Dropout layer~\cite{srivastava2014dropout}, which are respectively used to facilitate the training convergence and improve the model generalization capability.
We note that Batch Norm layers are not present in the original U-net architectures but, as also verified in preliminary experiments, their presence improves the segmentation performance.
The last encoder block (Enc Block 512 in Fig.~\ref{fig:unet}) does not have Dropout, while the last decoder block (Dec Block 32) is followed by a Conv layer with one $1 \times 1$ filter and a sigmoid activation.
Since the network is fully convolutional (there are no dense layers), it can process inputs of variable size.

We note that it would be possible to use more powerful FCN architectures such as DeepLabV3+~\cite{chen2018encoder}, which achieves state-of-the-art results in segmenting natural images.
However, models with a larger capacity, such as DeepLabV3+, require very large datasets to be trained on.
In remote sensing applications, a smaller network such as U-net is often preferred, given the limited amount of training data.
Moreover, the U-net outperforms other architectures in detecting small objects~\cite{Krestenitis2019b}, such as the snow avalanches in our work.

\subsection{Class balance}
Avalanches are small objects and the avalanche class is highly under-represented in the dataset (avalanche pixels are only $0.019\%$ of the total).
Therefore, a trivial model that classifies each pixel as ``non-avalanche'' would reach a classification accuracy of $99.98\%$.
A solution to handle class unbalance is to differently weight the loss relative to the pixels of the different classes so that the model is more penalized when it misclassifies the underrepresented class~\cite{kampffmeyer2016semantic}.
Specifically, we configured the loss to give twice more importance to the classification errors on the avalanche pixels.
We also experimented with loss functions specifically designed to handle class unbalance, such as the Jaccard-distance loss~\cite{csurka2013good} and the Lov{\'a}sz-Softmax loss~\cite{berman2018lovasz}, but we obtained worse results than optimizing the FCN using binary cross-entropy loss with class balancing.

\subsection{Data augmentation}
To avoid overfitting during training and to enhance the model generalization to new data, we perform data augmentation by randomly applying (on the fly) horizontal and vertical flips, horizontal and vertical shifts, rotations, zooming, and shearing to the training images. 
To ensure consistency, the same transformations on the input images are also applied to their labels (avalanche masks).

To compute the prediction of a whole SAR product at inference time, we could \textit{slide} the FCN on the large image and compute predictions for one window at a time.
However, this approach usually generates checkerboard artefacts and border effects close to the window edges.
To obtain smoother and more accurate predictions, we consider overlapping windows by sliding the FCN with a stride equal to half the window size.
Furthermore, we apply to each window all the possible 90$^{\circ}$ rotations and flips; then, we compute the predictions and, finally, revert the transformations on the predicted outputs.
To obtain the final segmentation, we first merge the multiple predictions available at each pixel location (stemming from the geometric transformations and the overlapping windows) and then we join them by using a $2^\text{nd}$ order spline interpolation.

\subsection{Attention mask}
\label{sec:attention}

\begin{figure*}[ht!]
 \centering
 \includegraphics[width=.85\textwidth]{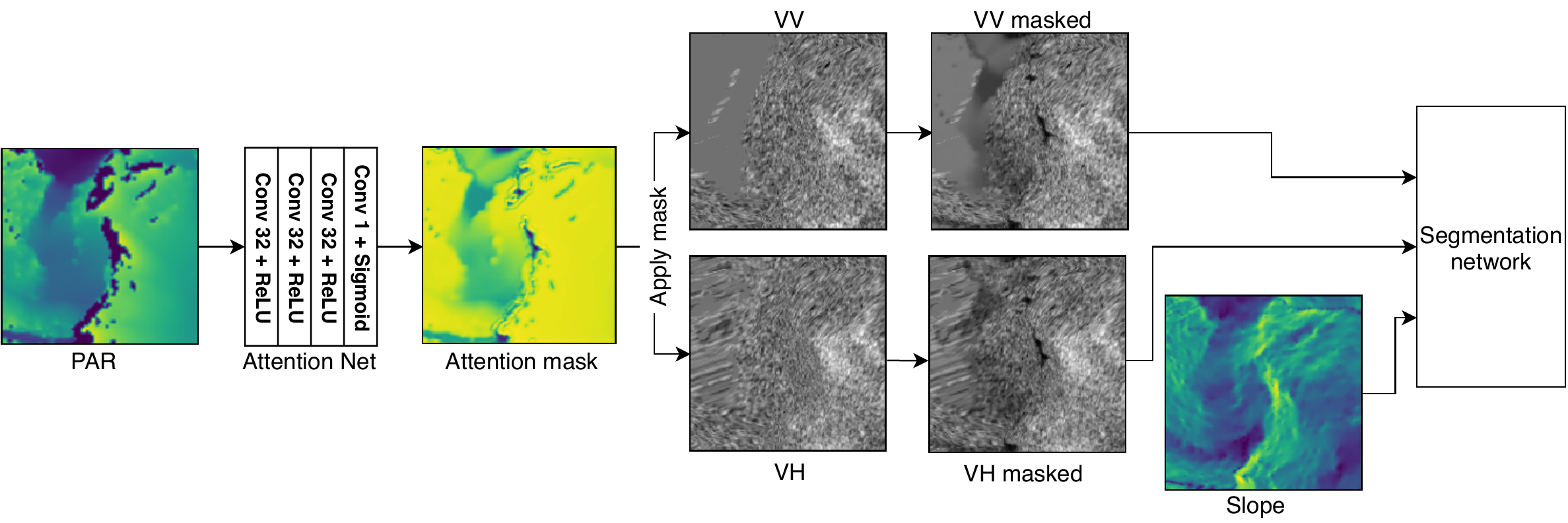}
 \vspace{-.4cm}
 \caption{For each patch, the Attention Net generates an attention mask from the PAR features and applies it to the VV, VH, and VVVH SAR features. 
 The masked SAR features and the slope (not masked) are then fed into the U-Net.
 Attention Net and U-net are jointly trained by minimizing the segmentation error. 
 Note that the VVVH feature is not shown in the figure for conciseness.}
 \label{fig:attention}
\end{figure*}

Following our hypothesis that the PAR feature map can highlight areas where it is more likely to find an avalanche, we propose a neural attention mechanism~\cite{xu2015show} that generates an \textit{attention mask} conditioned on the PAR.
The intention is to learn an attention mask that encourages the segmentation procedure to put more focus on specific regions of the input image.
Specifically, we use a small network that takes as input the PAR and generates the attention mask that is, subsequently, applied pixel-wise to the SAR channels (VV, VH, and VVVH) before they are fed into the segmentation network (see Fig.~\ref{fig:attention}).
We note that the attention mask is not applied to the input channel containing the slope feature map.

The attention network consists of three stacked Conv layers with 32 $3 \times 3$ filters and ReLU activations and a Conv layer with 1 $3\times3$ filter and sigmoid activation.
The attention network has a small receptive field (7 pixels), meaning that each attention value only depends on the local PAR. 
This is acceptable since the PAR already yields highly non-localized features from the DEM and captures long-range relationships in the scene.

The attention network is also fully convolutional and is jointly trained with the segmentation network. 
Our solution allows learning end-to-end on how to generate and apply the attention mask in a way that is optimal for the downstream segmentation task.
This is a more flexible approach than masking out parts of the input (e.g.\ by applying pre-computed runout masks), or directly pre-multiplying the SAR channels with the PAR feature map.

\subsection{Model training and evaluation}
We trained the FCN by feeding it with small square patches, rather than processing entire scenes at once, which would also be unfeasible due to the memory limitations of the GPU\footnote{Two Nvidia GTX2080 were used to train and evaluate the model}.
By using small patches it is also possible to inject stochasticity in the learning phase by randomly shuffling and augmenting the data at each epoch. 
This limits overfitting and decreases the chances of getting stuck in local minima.
We experimented with patches of $160 \times 160$ or $256 \times 256$ pixels, which is a size compatible with the receptive field of the filters in the innermost network layer (Enc Block 512), which is 140.
After preliminary experimentation, we obtained the best performance with the $160 \times 160$ patches.
The training and validation sets are generated by randomly partitioning these patches in order to prevent biasing either the training or validation sets towards any particular imaging parameters, such as the incidence angle. 
It should, moreover, be noted that image pairs are only constructed from the same satellite orbit number, such that the viewing geometries of the activity and reference images are nearly identical.
To build the training/validation set, we considered only the patches containing at least $1$ pixel classified as ``avalanche'' by the human expert.
We ended up with $\approx 35.000$ patches, of which $10\%$ were used as a validation set for model selection and early stopping.
Finally, out of the 118 available S1 scenes, one scene with date 17-Apr-2018, which contains 99 avalanches, was isolated from the rest and used as the test set.

\section{Results and discussion}

\begin{figure*}[!ht]
    \centering
    \subfigure{
        \includegraphics[width=0.7\textwidth]{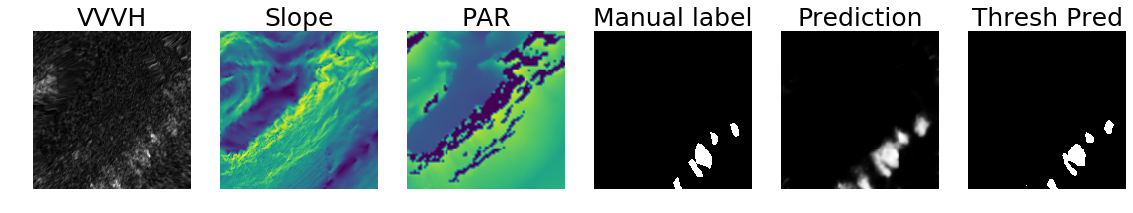}
    }

    \subfigure{
        \includegraphics[width=0.7\textwidth]{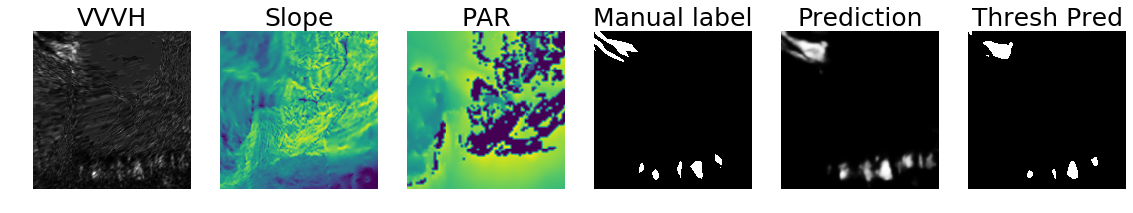}
    }

    \subfigure{
        \includegraphics[width=0.7\textwidth]{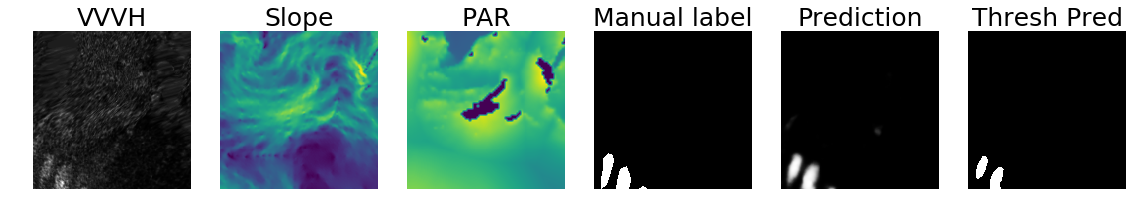}
    }

    \subfigure{
        \includegraphics[width=0.7\textwidth]{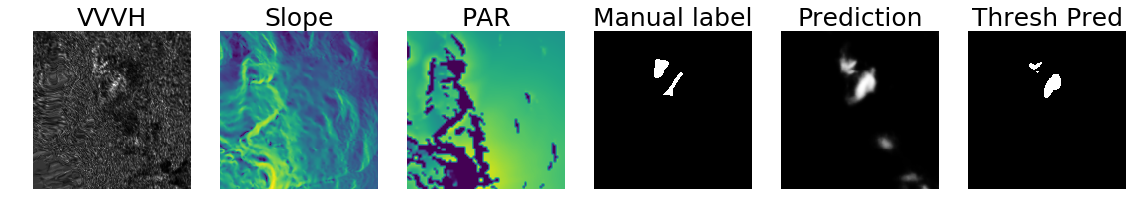}
    }
    \caption{Examples of prediction on individual patches of the validation set. From the left: i) VVVH input channel fed to FCN; ii) Slope feature fed to FCN; iii) PAR feature fed to Attention Net; iv) ground truth labels manually annotated by the expert; v) raw output of the FCN; vi) FCN output thresholded at 0.5.}
    \label{fig:patches_pred}
\end{figure*}

The network is trained with Adam optimizer~\cite{kingma2014adam} with default parameters; we used mini-batches of size 16 and dropout rate 0.4.
Examples of FCN predictions are depicted in Fig.~\ref{fig:patches_pred}. 
Since the networks predict real values in [0,1], a binary segmentation mask (last column) is obtained by thresholding the soft output (3\textsuperscript{rd} column) at 0.5.

Since the avalanche class is highly under-represented, accuracy is not a good measure to quantify the performance and, therefore, we evaluated the quality of the segmentation result by using different metrics.
The first is the F1 score, which is computed at the pixel level and is defined as
\[ F1 = 2 \frac{ \text{precision} \cdot \text{recall} }{ \text{precision} + \text{recall} },\]
where \textit{precision} is defined as $\frac{TP}{TP+FP}$ and \textit{recall} is $\frac{TP}{TP+FN}$ (TP = True Positives, FP = False Positives, FN = False Negatives).
The F1 score is also evaluated during training on the validation set and used for early stopping and for saving the best model.

To evaluate the segmentation results at a coarser resolution level, we considered the bounding boxes containing the avalanches in the ground truth and in the predicted mask.
To quantify how much the bounding boxes overlap in the ground truth and the predicted segmentation mask, we computed the intersection over union (IoU):
\[IoU = \frac{\text{Area of bounding boxes intersection}}{\text{Area of bounding boxes union}}.\]

We compared the proposed deep learning method with the state-of-the-art algorithm for automatic avalanche detection, which is currently used in production pipelines~\cite{vickerssynthetic}. 
Such a segmentation algorithm is primarily driven by change detection and filtering methods to enhance potential avalanche features; dynamic thresholding based on the statistics of image pairs controls the final delineated features. 
The baseline algorithm is, to a large extent, dependent on additional input layers such as slope, vegetation maps and runout zone information that restrict the areas where features are allowed to be detected, thereby reducing the number of false alarms as much as possible.

\begin{table}[ht!]
\setlength\tabcolsep{.9em} 
\small
\bgroup
\def\arraystretch{1.25} 
\centering
\caption{Segmentation results from the test image with 99 avalanches. We report the F1 score (in percent), Intersection over Union of the bounding boxes (in percent), True Positive (correct hits), False Negative (missed avalanches detection), and False Positive (false avalanches detection).}
\begin{tabular}{cccccc}
\cmidrule[1.5pt]{1-6}
\textbf{Method} & \textbf{F1 (\%)} & \textbf{IoU (\%)} & \textbf{TP (\#)} & \textbf{FN (\#)} & \textbf{FP (\#)} \\
\cmidrule[.5pt]{1-6}
Baseline & 38.13 & 33.11 & 44 & 45 & 11\\
FCN & 66.6 & 54.3 & 72 & 17 & 32 \\
\cmidrule[1.5pt]{1-6}
\end{tabular}
\label{tab:res}
\egroup
\end{table}

\begin{figure*}[ht!]
 \centering
 \includegraphics[width=\textwidth]{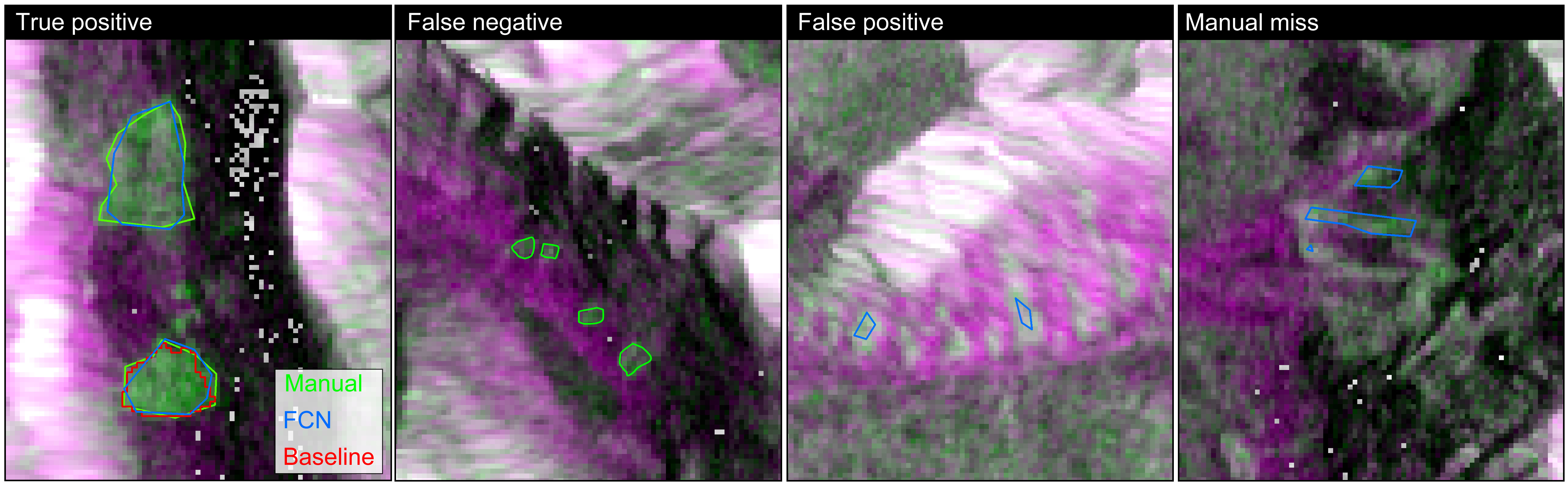}
 \vspace{-.8cm}
 \caption{Comparison between manual labeling and FCN output overlain onto a RGB change detection image. From the left: i) agreement between FCN detection and manual annotations; ii) avalanches missed by the FCN; iii) false detection from the FCN algorithm; iv) avalanches correctly detected by the FCN but overlooked during the manual annotation.}
 \label{fig:visual_eval}
\end{figure*}

Tab.~\ref{tab:res} reports the results obtained on the test image.
Compared to the baseline, the FCN achieved a much higher agreement with the manual labels, as indicated by the higher F1 and IoU values.
Out of the 99 avalanches in the test image, FCN correctly identified 72 of them and missed 17.
However, most of the FN are small avalanches that are difficult to detect.
FCN also identified 32 FP: most of them are due to particular terrain structures, which cause high backscatter that resembles avalanches (see Fig.~\ref{fig:visual_eval}).
Interestingly, some of those FPs are actual avalanches that have been overlooked during the manual annotation. 

\subsection{Ablation study}
The ablation study consists of removing some features from the model or from the input data to evaluate how these affect the performance.
In particular, we study how much each SAR channel and the topographical feature maps contribute to the segmentation results.
We also evaluate the difference in concatenating the PAR to the other input channels (VV, VH, VVVH, and slope) or using it to compute the attention mask that is applied pixel-wise to the SAR channels (see the details in Sect.~\ref{sec:attention}).

The results reported in Tab.~\ref{tab:abl_res} indicate that the most important improvement comes from including the difference image obtained by the VH channels, compared to using the VV channel alone.
By adding the slope and PAR features it is possible to further increase the segmentation performance.
Finally, the results show that the proposed attention mechanism allows to better exploit the information yield by the PAR, compared to just concatenating the PAR feature map to the other input channels.

\begin{table}
\setlength\tabcolsep{.65em} 
\small
\bgroup
\def\arraystretch{.9} 
\centering
\caption{Ablation experiment results.}
\begin{tabular}{ccccccc}
\cmidrule[1.5pt]{1-7}
\textbf{VV} & \textbf{VH} & \textbf{VVVH} & \textbf{Slope} & \textbf{PAR} & \textbf{PAR (attn.)} & \textbf{F1} \\
\cmidrule[.5pt]{1-7}
\cmark & & & & & & 55.4 \\
\cmark & \cmark & & & & & 63.0 \\
\cmark & \cmark & \cmark & & & & 64.9 \\
\cmark & \cmark & \cmark & \cmark & & & 65.2 \\
\cmark & \cmark & \cmark &  \cmark & \cmark & & 65.4 \\
\cmark & \cmark & \cmark & \cmark & & \cmark & 66.6 \\
\cmidrule[1.5pt]{1-7}
\end{tabular}
\label{tab:abl_res}
\egroup
\end{table}

\section{Conclusions}

In this work, we proposed the first deep learning approach for saliency segmentation of avalanches in Sentinel-1 SAR images. 
As channels of the images provided as input to the segmentation network, we used the time difference of the radar backscatter information, as well as topographical information.
The latter consists of the terrain slope and the newly introduced potential angle of reach, which indicates the likelihood of finding avalanches at different locations.
The topographical feature maps were provided along with the SAR features to a Fully Convolutional Network, which was trained to perform avalanche segmentation.
The ground truth segmentation masks used to train the deep learning model came from the manual labelling of avalanche pixels performed by a human expert. 
A total of 118 Sentinel-1 SAR products were labelled, of which 117 were used for training and one single product was used for testing the segmentation performance on unseen data.

The Fully Convolutional Network was extended with an additional attention block, jointly trained with the rest of the segmentation network, which computes an attention mask conditioned on the potential angle of reach.
The mask was applied to the input SAR features to let the segmentation network focusing more on the critical areas.

The results show the effectiveness of the proposed method, improving the F1 score of 38.1\% achieved by a baseline signal processing algorithm to 66.6\%.
The F1 score was computed based on the manual labelling of the human expert. 
The proposed deep learning model only fails to detect some of the smaller avalanches, while detects additional avalanches that have been missed by the expert. 

By being the first of its kind, we believe that our work will pave the way for pixel-level classification of snow avalanches in SAR data with deep learning and will serve as a future reference in the field of Earth science and remote sensing.
Our analysis and the obtained results suggest that the potential angle of reach is well correlated with the presence of avalanches.
Therefore, we believe that the proposed potential angle of reach feature will be useful for future work in this field.
In the next step, we aim to extend our dataset to evaluate the FCN's performance on SAR images with different snow conditions (wet or dry).

\bibliographystyle{IEEEtran}
\bibliography{references}

\end{document}